\documentclass[12pt]{article}

\usepackage{times, xcolor, graphicx, graphics, amsmath, amssymb, booktabs, multicol,doi}
\usepackage[pagewise, modulo]{lineno}
\usepackage{hyperref}
\hypersetup{
	colorlinks,
	citecolor=black,
	filecolor=black,
	linkcolor=black,
	urlcolor=black
}
\usepackage[labelfont={bf}]{caption}
\usepackage[backend=bibtex, bibstyle=ieee, citestyle=numeric-comp, sorting=none, doi=true, isbn=false, url=false, maxbibnames=99]{biblatex}

\newcounter{lastnote}

\begin{document} 
	\sloppy

	\baselineskip24pt

	{\parindent0pt 
		
		\Huge{Deformation localisation in ion-irradiated FeCr} 
		
		\bigskip
		
		\Large
		{Kay Song $^{\text{a}, \star}$, Dina Sheyfer $^{\text{b}}$, Wenjun Liu $^{\text{b}}$, Jonathan Z Tischler $^{\text{b}}$, Suchandrima Das $^{\text{c}}$, Kenichiro Mizohata $^{\text{d}}$, Hongbing Yu $^{\text{f}}$, David E J Armstrong $^{\text{e}}$, Felix Hofmann $^{\text{a}, \dagger}$}
		
		\bigskip
		\large{$^{\text{a}}$ Department of Engineering Science, University of Oxford, Parks Road, Oxford, OX1 3PJ, UK} \\
		\large{$^{\text{b}}$ X-ray Science Division, Argonne National Laboratory, 9700 South Cass Avenue, Lemon, IL 60439, USA} \\
		\large{$^{\text{c}}$ Department of Mechanical Engineering, University of Bristol, Queen's Building, University Walk, Bristol, BS8 1TR, UK} \\
		\large{$^{\text{d}}$ Department of Physics, University of Helsinki, P.O. Box 43, 00014 Helsinki, Finland} \\
		\large{$^{\text{e}}$ Department of Materials, University of Oxford, Parks Road, Oxford, OX1 3PH, UK}\\
		\large{$^{\text{f}}$ Canadian Nuclear Laboratories, Chalk River, ON K0J 1J0, Canada} \\
		
		\bigskip
		\large{$^{\star}$Corresponding author email: kay.song@eng.ox.ac.uk} \\
		\large{$^{\dagger}$felix.hofmann@eng.ox.ac.uk}
		
		\bigskip
		\begin{multicols}{2}
		\large{ORCID:}\\
		\normalsize{Kay Song: 0000-0001-8011-3862 \\ Dina Sheyfer: 0000-0003-4189-3899 \\ Wenjun Liu: 0000-0001-9072-5379 \\ Jonathan Z Tischler: 0000-0001-6163-7505 \\ Suchandrima Das: 0000-0002-3013-219X \\ \\ Kenichiro Mizohata: 0000-0003-1703-2247 \\ Hongbing Yu: 0000-0002-6527-9677 \\ David E J Armstrong: 0000-0002-5067-5108 \\ Felix Hofmann: 0000-0001-6111-339X}
		\end{multicols}
	}
	\newpage
	\begin{abstract}
		Irradiation-induced ductility loss is a major concern facing structural steels in next-generation nuclear reactors. Currently, the mechanisms for this are unclear but crucial to address for the design of reactor components. Here, the deformation characteristics around nanoindents in Fe and Fe10Cr irradiated with Fe ions to $\sim$1 displacement-per-atom at 313 K are non-destructively studied. Deformation localisation in the irradiated materials is evident from the increased pile-up height and slip step formation, measured by atomic force microscopy. From 3D X-ray Laue diffraction, measurements of lattice rotation and strain fields near the indent site show a large confinement, over 85\%, of plasticity in the irradiated material. We find that despite causing increased irradiation hardening, Cr content has little effect on the irradiation-induced changes in pile-up topography and deformation fields. The results demonstrate that varying Cr content in steels has limited impact on retaining strain hardening capacity and reducing irradiation-induced embrittlement.
		
	\end{abstract}
	
	\medskip
	
	\begin{center}
		\small{\textbf{Graphical Abstract}}
		\begin{figure}[h!]
			\includegraphics[width=0.9\textwidth]{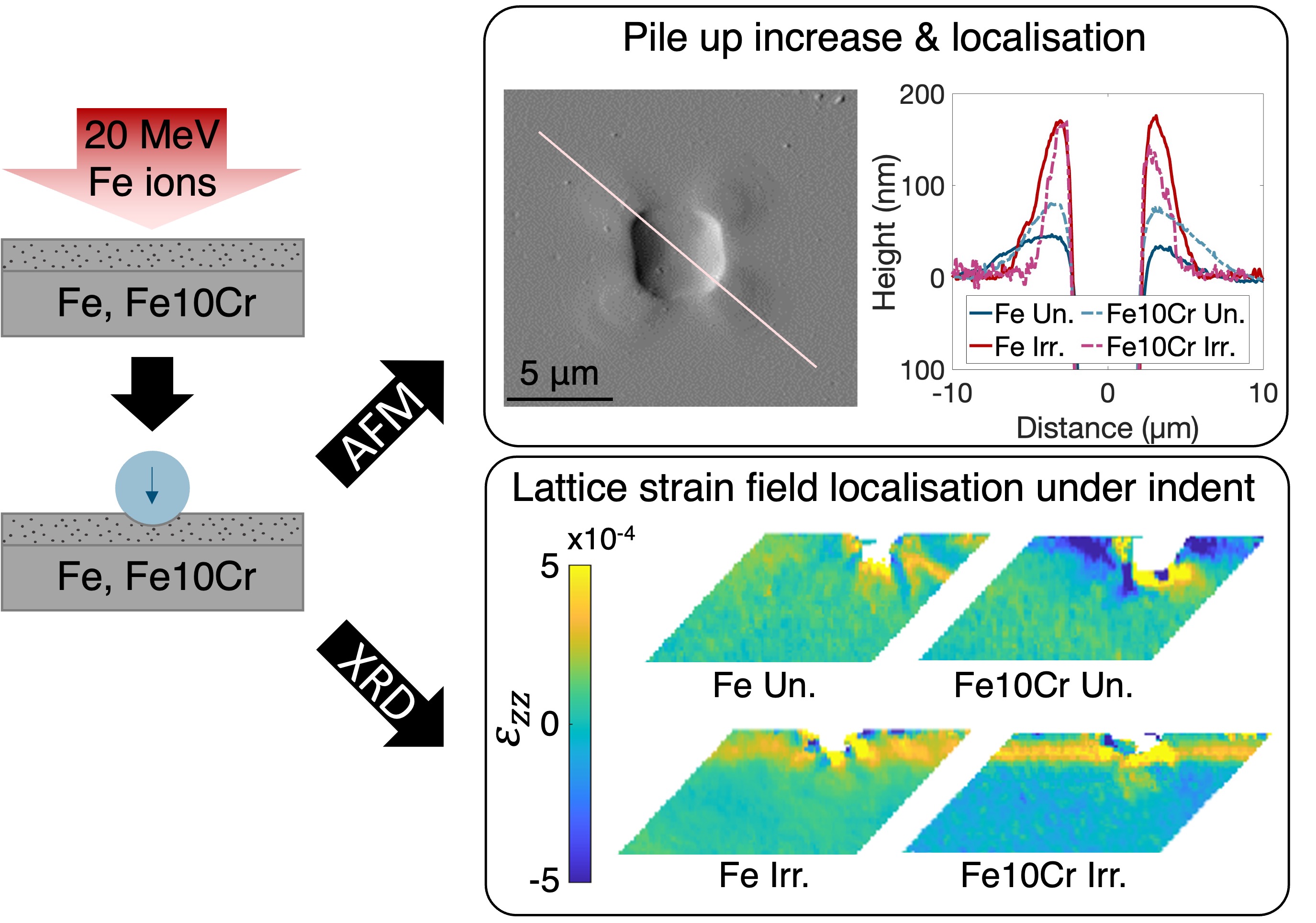}
			\centering
		\end{figure}
		
	\end{center}
	
	\newpage
	
		
		\section*{Main Text}
		Understanding the mechanical behaviour of structural materials is essential for the design and operation of next-generation nuclear reactors. Reduced-activation ferritic/martensitic steels are promising candidates due to their resistance to irradiation swelling and good thermomechanical properties \cite{Baluc2004}. Nevertheless, neutron irradiation in reactor environments will still lead to increased hardening and reduced ductility of structural steels \cite{Kurtz2019, Zinkle2005}. Studies into the effect of microstructure and composition to mitigate these irradiation-induced changes are crucial \cite{Bhattacharya2021, Rieth2021, Oh2009}.
		
		A well-known phenomenon following the deformation of irradiated materials is deformation localisation \cite{Sharp1967, Farrell2004, Hashimoto2004, Byun2006a, Zinkle2012, Das2019b}. The accepted mechanism is that glide dislocations annihilate irradiation-induced defects \cite{Saada1962, Foreman1969}. This creates easy glide channels with a locally reduced density of irradiation-induced defects that confine subsequent deformation \cite{Little1976}. The formation of these channels contributes to irradiation embrittlement and irradiation-assisted stress corrosion cracking \cite{Zinkle2013}. 
		
		Presently, the exact details of the mechanisms by which irradiation defects are removed still remain unclear. This is an active area for modelling and simulations \cite{Patra2013, Patra2016, Cui2021}. However, there is a lack of experimental investigations to validate modelling predictions. TEM is a common way to directly observe dislocation microstructures and channels \cite{Zinkle2006, Gussev2015, Dai2001}. However, TEM is unable to provide information, especially in 3D, on the local stress and strain states, which are key to compare with simulations. 
		A non-destructive method such as X-ray diffraction can uniquely provide information over a greater volume and in 3D \cite{Das2018, Altinkurt2018}.  
		
		Here, we consider the deformation microstructure around nanoindents in Fe and Fe10Cr for both the unirradiated material and following irradiation with Fe ions at 313 K. This allows the study of the effect of irradiation and Cr content. Surface topography around the indent site was examined with atomic force microscopy (AFM). 3D lattice rotation and strain fields under the surface were probed by X-ray Laue diffraction. The samples used in this study are the same as those in previous studies of irradiation-induced lattice strain, thermal and mechanical properties \cite{Song2020, Song2022, Song2023}. The materials from which the samples are taken have also been investigated extensively in other irradiation studies \cite{Schaublin2017, Hardie2013a, Bhattacharya2016, Brimbal2014, Brimbal2013, Roldan2016, Lin2021}. This is important, as it helps to build a comprehensive understanding of the different aspects of irradiation damage whilst maintaining consistent sample processing and irradiation history \cite{Xiao2020}.

		
		The Fe and Fe10Cr materials were manufactured under the European Fusion Development Agreement (EFDA) programme (contract EFDA-06-1901) and their chemical composition is listed in Table \ref{tab:comp}. The materials were made by induction melting under an argon atmosphere followed by cold-forging and heat treatments \cite{Coze2007, Fraczkiewicz2011}. 

		\begin{table} [h!]
			\centering
			\begin{tabular}{ ccccccc }
				\toprule
				Alloy	&	Cr (wt \%)	&	C (wppm)	&	S (wppm)	&	O (wppm)	&	N (wppm) \\
				\midrule
				Fe	&	$<$0.0002	&	4	&	2	&	4	&	1	\\
				\addlinespace
				Fe10Cr	&	10.10	&	4	&	4	&	4	&	3	\\
				\bottomrule
			\end{tabular}	
			\caption{Manufacturer-provided chemical compositions as measured by glow discharge mass spectrometry \cite{Coze2007, Fraczkiewicz2011}.}
			\label{tab:comp}	
		\end{table}
		
		The samples were mechanically ground with SiC paper, and then polished with diamond suspension and colloidal silica (0.04 \textmu m). Finally, they were electropolished to achieve a deformation-free surface finish.
		
\begin{figure}[h!]
			\centering
			\includegraphics[width=\textwidth]{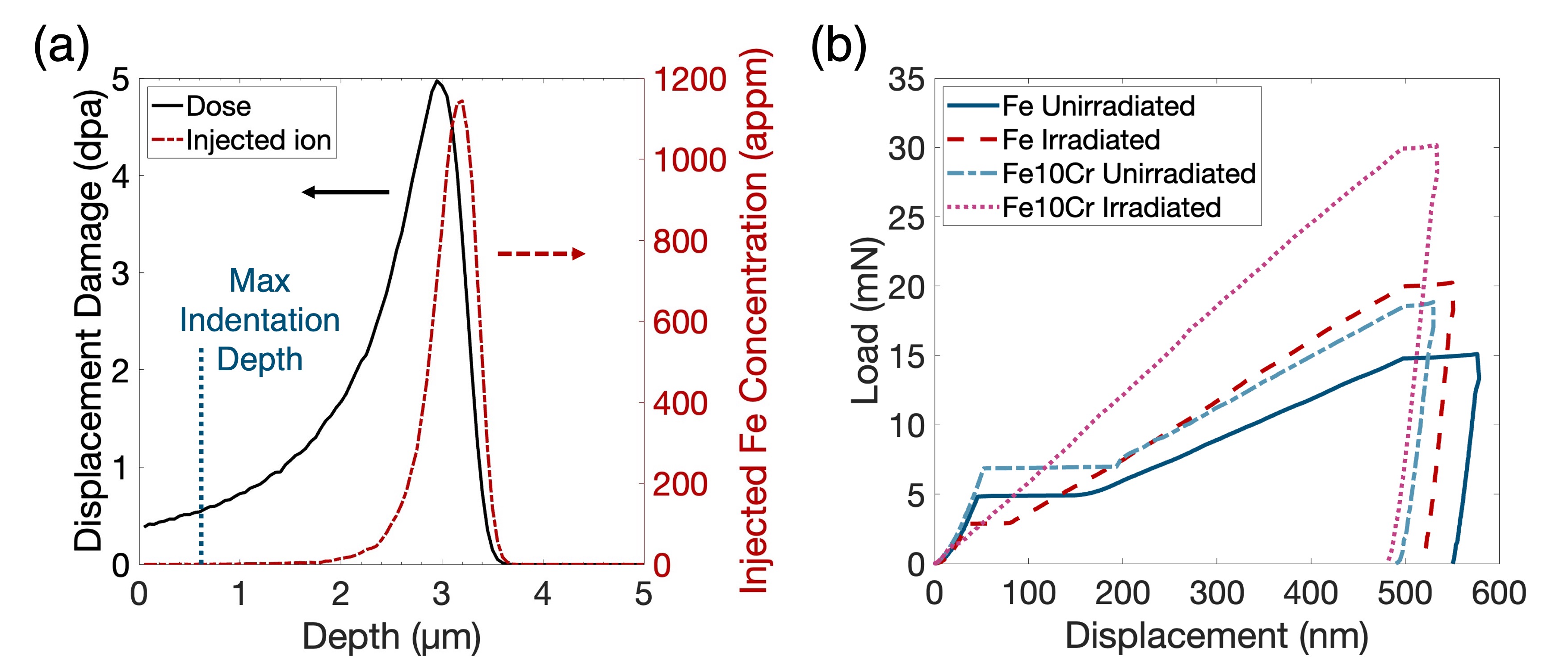}
			\caption{(a) The irradiation damage and injected ion profiles as calculated by SRIM \cite{Ziegler2010}. (b) Representative load-displacement curves from each material.}
			\label{fig:dpa_ld}
		\end{figure}

		Ion irradiation was performed with 20 MeV Fe$^{4+}$ ions using the tandem accelerator at the University of Helsinki. The samples were actively held at 313 K during irradiation. The dose profile shown in Figure \ref{fig:dpa_ld}(a) was calculated with SRIM using the Quick K-P model \cite{Ziegler2010} with 20 MeV Fe ions on a Fe target with 40 eV displacement energy \cite{ASTME5212016} at normal incidence. The calculated damaged layer extends to 3.5 \textmu m below the sample surface. The average dose in the top 580 nm, which is the maximum indentation depth (including creep), is 0.46 displacements-per-atom (dpa). The damage profile peaks at 5 dpa at a depth of 3 \textmu m. The average dose of the whole damage layer is 1.7 dpa. 
		
		Consistent grain orientation across different samples is important as the pile-up morphology is expected to show orientation dependence from previous studies of body-centred cubic (BCC) materials \cite{Das2019, Das2020}. Those studies also found that pile-up heights are the most sensitive to irradiation-induced changes when indenting along the $\langle 001 \rangle$ direction. In this study, electron backscatter diffraction was used to identify grains within 5$^{\circ}$ of $\langle 001 \rangle$ out-of-plane orientation on each sample for investigation.
		
		Nanoindentation was performed using an MTS Nano Indenter XP with a spherical diamond tip of 5 \textmu m nominal radius (Synton-MDP). Load-controlled indentation was performed to a nominal depth of 500 nm, with a hold segment of 10 seconds before unloading. The indentation depth was chosen to be less than 20\% of the damaged layer thickness to ensure that a majority of the plastic zone is contained in the irradiated layer \cite{Mata2006, Chen2006}.
		
		The load-displacement curves (Figure \ref{fig:dpa_ld}(b)) for all materials show similar initial Hertzian elastic contact segments. 
		Pop-in events occurred in all samples, and the irradiated samples show lower pop-in loads than the unirradiated samples possibly due to irradiation defects acting as extra sources of dislocation nucleation \cite{Song2022, Jin2018, Pathak2017}.
		
		Analysis of the peak loads of at least 6 indents on each sample was performed to obtain an indication of the material hardness. Large pop-in loads were often observed for the unirradiated materials due to low populations of pre-existing dislocations. As a result, many indentation tests on the pre-identified grains exceed 500 nm in depth, and could not be used. 
		
		Solid solution hardening in the unirradiated materials is observed from a 20\% ($\pm 6$\%) increase in peak load between Fe and Fe10Cr. Irradiation resulted in a 21\% ($\pm 8$\%) increase in peak load for Fe and a 43\% ($\pm 7$\%) increase for Fe10Cr. This agrees with hardness trends from previous measurements with Berkovich nanoindentation, which show a 31\% ($\pm 5$\%) increase for Fe and a 48\% ($\pm 7$\%) increase for Fe10Cr following irradiation \cite{Song2023}. The creep distance from holding the peak load is greater in Fe than Fe10Cr for both the unirradiated and irradiated materials.
		
\begin{figure}[h!]
			\centering
			\includegraphics[width=0.95\textwidth]{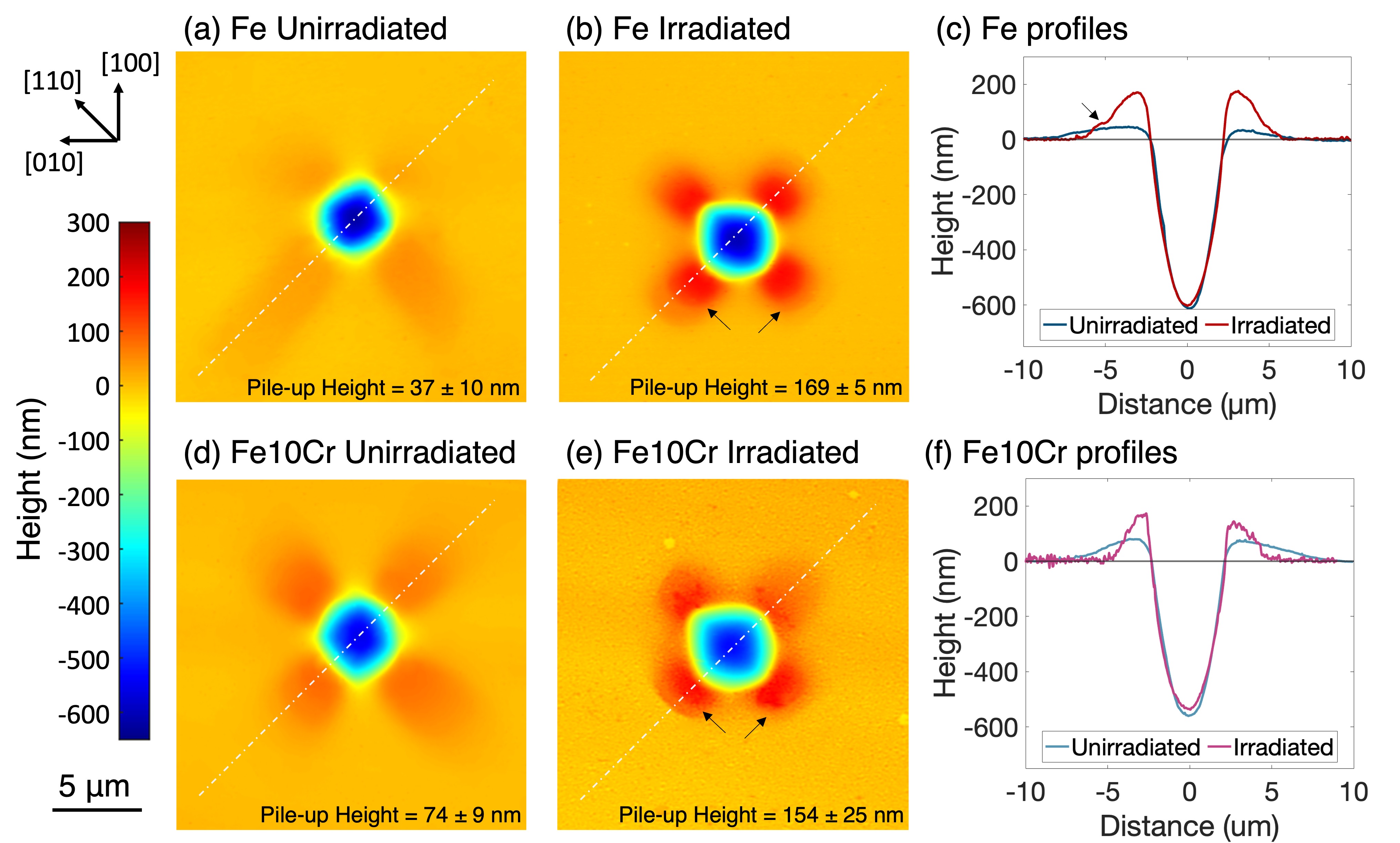}
			\caption{The pile-up topography surrounding the indent site for (a) unirradiated Fe and (b) irradiated Fe. The average pile-up heights are calculated from repeated measurements of the 4 pile-up lobes surrounding each indent, with the associated standard deviations shown. The white dashed line indicates the profile of pile-up height taken from the measurements and shown in (c). The same layout for Fe10Cr is shown in (d)--(f). Black arrows indicate evidence of slip steps (further details in the Supplementary File).}
			\label{fig:afm}
		\end{figure}

		Pile-up topography around the indent sites (Figure \ref{fig:afm}) was measured with a Veeco MultiMode 8 atomic force microscope (AFM) in ScanAsyst\textsuperscript{TM} mode. There is a four-fold symmetry in the pile-up pattern due to the indents being made on grains with close to $\langle 001 \rangle$ out-of-plane orientation. The pile-up lobes are oriented along the $\langle110\rangle$ directions, which has been previously observed in other BCC materials \cite{Das2019, Das2020, Yu2020, Smith2003}.
		
		For the unirradiated samples, the pile-up height increases with Cr content from $37 \pm 10$ nm for Fe to $74 \pm 9$ nm for Fe10Cr. Pile-up has been found to increase with decreasing strain hardening exponents \cite{Taljat2004, Hill1989}. This suggests that the presence of Cr reduces strain hardening capacity. Indeed, this is consistent with our previous study of the same materials \cite{Song2022}, which shows from nanoindentation stress-strain curves that strain hardening capacity decreases with Cr content in unirradiated FeCr.
		
		Pile-up height increases after irradiation in both Fe and Fe10Cr to $169 \pm 5$ nm and $154 \pm 25$ nm, respectively. 
		The increase in average pile-up height is consistent with previous findings of irradiation-induced strain hardening capacity reduction \cite{Song2022, Qu2020, Xiao2019}. Cr content has little effect on the pile-up height following irradiation. This is consistent with previous findings from stress-strain studies of the same materials, where strain hardening exponents were reduced to zero regardless of Cr content, even for a dose that is 90\% lower than the current study \cite{Song2022}. Hardie \textit{et al.} \cite{Hardie2015} also observed an increase in pile-up height for Fe12Cr following irradiation to 6.18 dpa. However, the orientation of the grains was not specified, which is known to significantly affect pile-up height \cite{Das2019, Das2020, Zayachuk2017}. 
		
		The pile-up topography is more localised in the irradiated samples, only extending to $\pm$ 5 \textmu m on either side of the indent centre along the $\langle110\rangle$ direction (Figure \ref{fig:afm}(c) and (f)). In contrast, the lobes of the unirradiated materials extend up to 10 \textmu m away from the indent. Slip steps are also observed in the pile-up of the irradiated material (black arrows in Figure \ref{fig:afm}(b), (c), and (e)). Hardie \textit{et al.} \cite{Hardie2013} previously found that slip steps on the surface of ion-irradiated Fe12Cr result from shear bands beneath the surface, associated with defect-free channels in the irradiated layer. These channels lead to irradiation-induced strain softening and localisation of deformation \cite{Singh1999}.  
		
		
		
		Lattice rotation and strains around the nanoindents were measured with micro-beam X-ray Laue diffraction at the 34-ID-E beamline at the Advanced Photon Source (Argonne National Laboratory, IL, USA). Differential Aperture X-ray Microscopy (DAXM) was used to obtain depth resolution into the surface of the material. The technique has been described in detail elsewhere \cite{Larson2002, Liu2004, Das2018a}. 
		Here, the sample was mounted in a 45$^{\circ}$ reflection geometry. The polychromatic incident beam (7--30 keV) is focused on the sample surface with a size of $170 \times 240$ nm$^{2}$. For each sample, a $30 \times 40 \times 20$ \textmu m$^{3}$ volume was probed with $1.1 \times 1.5 \times 1$ \textmu m$^{3}$ 3D spatial resolution. Due to the sample mounting geometry, the path of the beam was at 45$^{\circ}$ to the sample surface. Hence the edges of the volume probed by X-rays is also confined by that angle (Figure \ref{fig:rot} and \ref{fig:strain}). Further details are included in the Supplementary File.
		
		The orientation distribution and strain refinement calculations \cite{Chung1999} were performed using LaueGo \cite{Tischler2020a}. The calculation of lattice rotation from the orientation of each voxel, data interpolation, and visualisation of the lattice rotation and strain field were performed with code modified from \cite{Das2018, Das2019b}. The reference lattice orientation was calculated from the average orientation between 18--20 \textmu m below the sample surface.

		\begin{figure}[h!]
			\centering
			\includegraphics[width=0.95\textwidth]{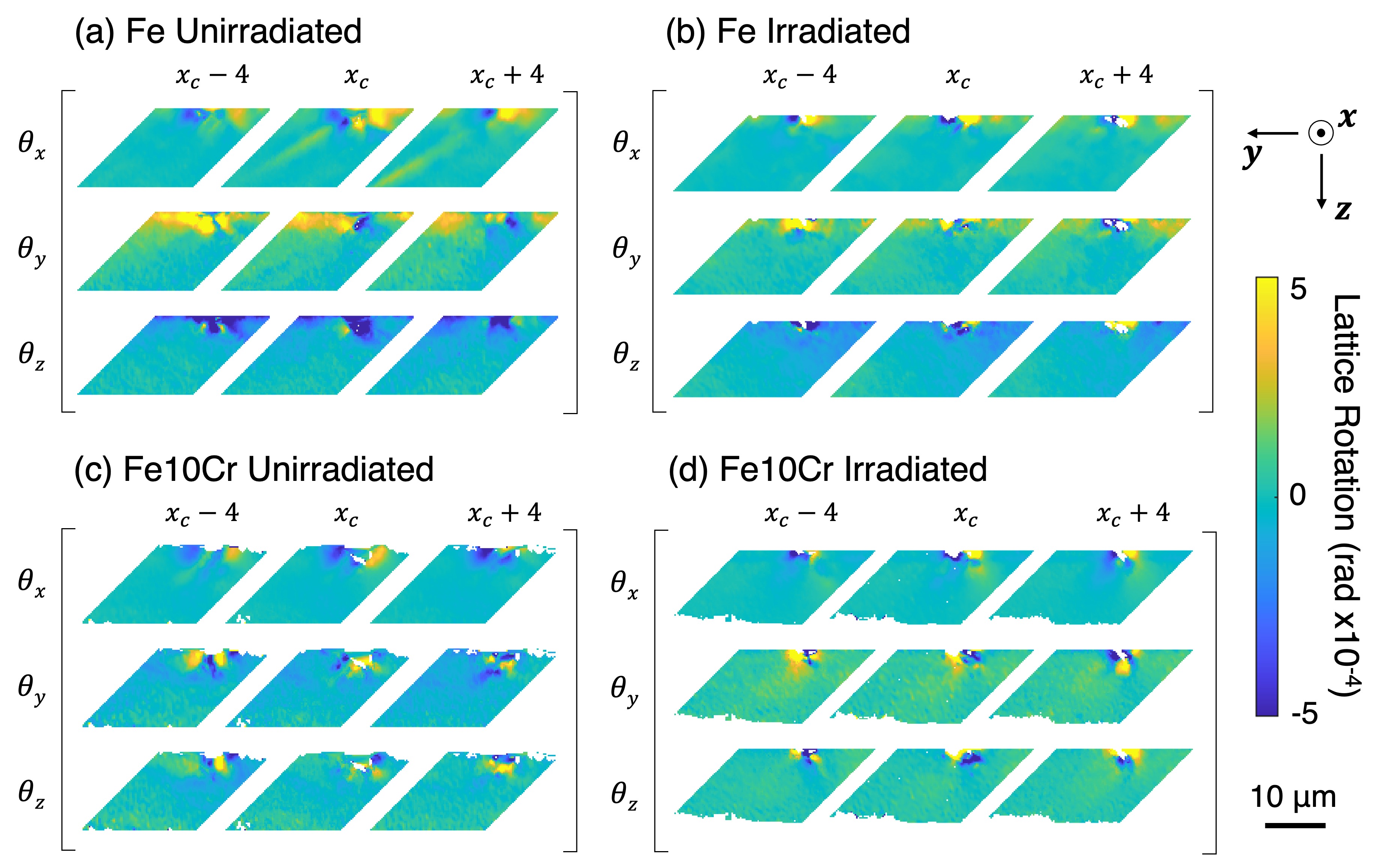}
			\caption{The lattice rotation under the indent for the different samples in this study. The $\boldsymbol{x}-\boldsymbol{y}$ plane is parallel to the surface of the bulk sample, while $\boldsymbol{z}$ points into the bulk material. For each sample, the lattice rotations components ($\theta_{x}$, $\theta_{y}$, and $\theta_{z}$) are presented from 3 slices (parallel to the $\boldsymbol{y}-\boldsymbol{z}$ plane). The middle slice ($x = x_{c}$) is taken through the centre of the indent. The slices on either side are at a distance of 4 \textmu m from the centre in the $\boldsymbol{x}$-direction. }
			\label{fig:rot}
		\end{figure}
		
		
		The lattice rotations in the unirradiated Fe extend to over 20 \textmu m away from the indentation site (Figure \ref{fig:rot}(a)). There is a long `streak' present in the $\theta_{x}$ map, similar to previous measurements in tungsten \cite{Das2018}. The rotation fields of Fe10Cr (Figure \ref{fig:rot}(c)) are similar in spatial extent and distribution to Fe.
		
		The rotation field induced by indentation in the irradiated samples is more localised than in the unirradiated material. This is particularly noticeable for the Fe sample (Figure \ref{fig:rot}(b)), where the rotation field is confined to the top 5 \textmu m in the $\boldsymbol{z}$-direction, and $\sim$7 \textmu m in the $\boldsymbol{x}$- and $\boldsymbol{y}$-directions. The distribution of lattice rotations in the irradiated Fe10Cr sample (Figure \ref{fig:rot}(d)) is similar to the irradiated Fe, with some extended regions of smaller lattice rotations ($|\theta| < 3\times 10^{-4}$ radians) up to 10 \textmu m from the indent centre. This represents a reduction of over 87\% of the volume containing lattice rotations.
		

		\begin{figure}[h!]
			\centering
			\includegraphics[width=0.95\textwidth]{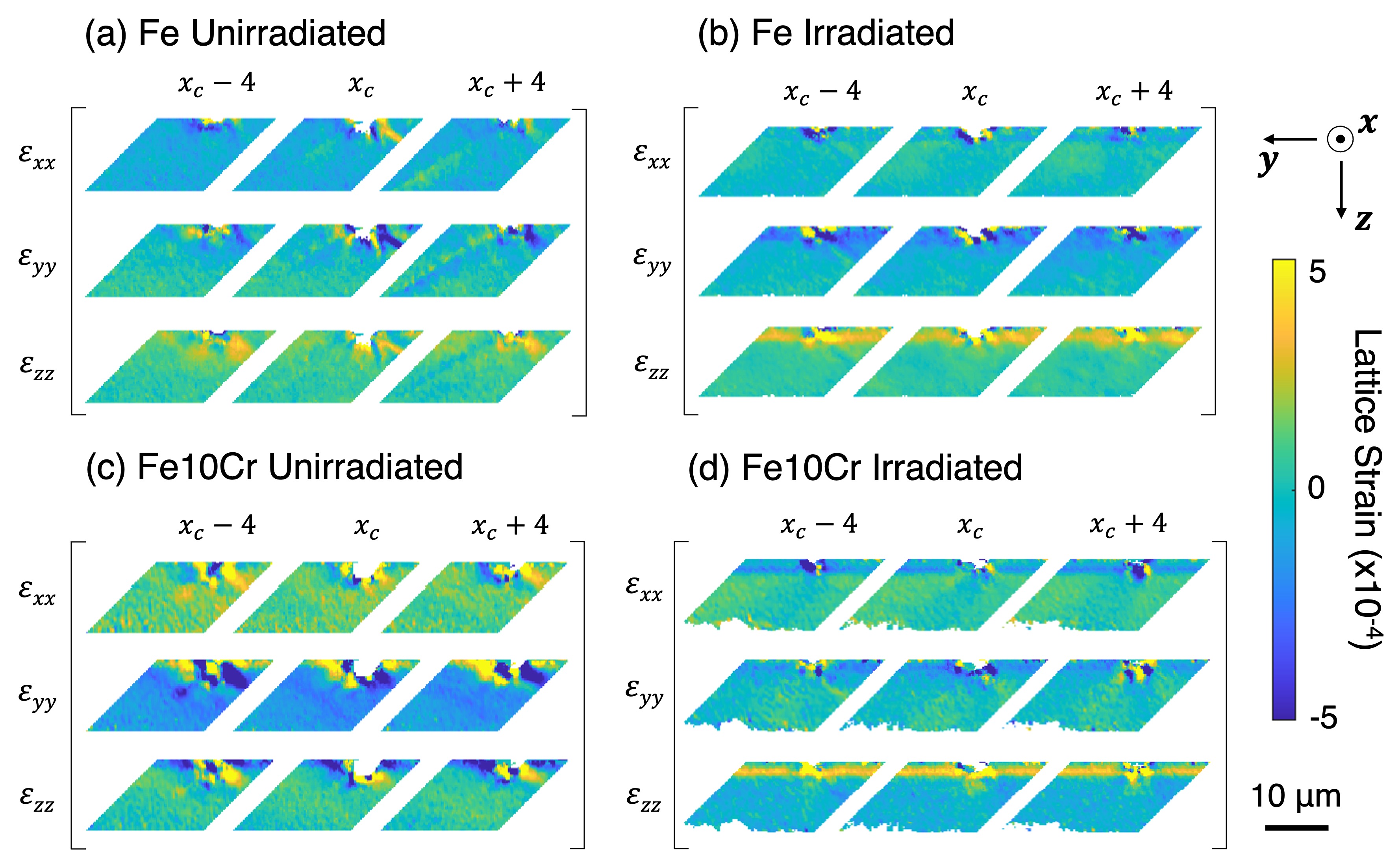}
			\caption{The 3 principal components of the deviatoric strain tensor, $\epsilon_{xx}$, $\epsilon_{yy}$, and $\epsilon_{zz}$ in the samples. The orientation of the measurements and the slicing convention are the same as described in Figure \ref{fig:rot}. }
			\label{fig:strain}
		\end{figure}
		
		
		The spatial distribution of tensile (positive) and compressive (negative) strain is similar between the unirradiated samples of Fe and Fe10Cr (Figure \ref{fig:strain}). However, the spatial extent and magnitude of strain, both tensile and compressive, are greater in Fe10Cr than in Fe. This could be attributed to the greater peak load applied to Fe10Cr to reach the same indentation depth as in Fe (Figure \ref{fig:dpa_ld}(b)), which in turn leads to greater residual stresses.
		
		After irradiation, there is a large tensile strain in the $\boldsymbol{z}$ direction for Fe and Fe10Cr at a depth of $\sim$3 \textmu m (Figure \ref{fig:strain}(b) and (d)). This is likely due to the presence of injected Fe ions from irradiation, as previously observed \cite{Song2020, Song2023}. There is a corresponding compressive strain of lower magnitude in the $\boldsymbol{x}$- and $\boldsymbol{y}$-directions. Furthermore, the magnitudes of $\epsilon_{xx}$ and $\epsilon_{yy}$ are similar, which is consistent with irradiation-induced lattice swelling and the presence of the external boundary condition requiring geometric continuity between the irradiated layer and the unirradiated bulk \cite{Das2018a}. 
		
		For both the irradiated Fe and Fe10Cr materials, the spatial extent of the strain field beneath and surrounding the indent is mostly restricted in the $\boldsymbol{z}$-direction to the top 3--4 \textmu m (Figure \ref{fig:strain}(b) and (d)). The containment of the strain field is largely within the irradiated layer, with rapidly vanishing strains at depths beyond that. Similar to lattice rotation fields, the extent of the strain fields in the $\boldsymbol{x}$- and $\boldsymbol{y}$-directions is also reduced following irradiation for both Fe and Fe10Cr. It appears that Cr content does not significantly affect the strain states following irradiation and deformation. This is despite a $\sim$50\% increase in applied load for the irradiated Fe10Cr material to achieve the same indentation depth as the irradiated Fe sample.
		
		The localisation of lattice rotation and strain fields in the irradiated Fe and Fe10Cr is consistent with previous observations of plastic zone confinement in irradiated Fe12Cr \cite{Hardie2013, Hardie2015}. From TEM lift-outs taken through indent cross-sections, it was observed that the spatial extent of dislocation propagation reduced by 24--31\% in the irradiated material.
		
		
		Combining the nanoindentation, AFM, and DAXM measurements, we can formulate a comprehensive picture of plasticity in ion-irradiated Fe and Fe10Cr. Without irradiation, the presence of Cr causes solid solution hardening. This necessitates a greater indentation load for Fe10Cr than Fe to reach the same indentation depth. As such, the residual stresses and strains are greater. Furthermore, the presence of Cr also reduces the strain hardening capacity of the material, which leads to a greater pile-up height in Fe10Cr.
		
		Ion irradiation introduces defects in the material that act as obstacles to glide dislocations, causing hardening. Even though the inclusion of 10\%Cr causes a similar amount of net hardening as introducing irradiation defects in Fe, the respective pile-up topography and residual lattice rotation and strain fields are very different. 
		The great increase in pile-up height due to irradiation ($\sim$450\% for Fe) reflects a dramatic reduction in strain hardening capacity, as previously observed from nanoindentation stress-strain curves of the same materials \cite{Song2022}. In comparison, the pile-up height only increases $\sim$200\% from the addition of 10\%Cr to Fe in the unirradiated case, despite showing similar hardening. The mechanism of strain softening in the irradiated material can be explained by the observation of slip steps, an indication of dislocation channelling, which provides paths for subsequently-generated glide dislocations to move at a lower stress. Irradiation-induced strain softening also causes a corresponding localisation of pile-up topography, further indicating deformation localisation. 
		We directly observed plastic zone confinement in the irradiated materials from the localisation of lattice strain and rotation fields surrounding the indents. 
		
		The effect of Cr on the post-irradiation deformation behaviour is limited. Even though the irradiation-induced hardening is much greater in Fe10Cr than in Fe, the height and spatial extent of the pile-up topography are very similar for both irradiated materials. This supports previous observations that post-irradiation reduction in strain hardening capacity does not depend on Cr \cite{Song2022}. In this study, we also observed the effect of this on the lattice rotation and strain fields in both Fe and Fe10Cr, which were reduced to similar extents following irradiation. The 3D measurements in this study also provide valuable comparisons for future modelling studies, which will be crucial to understanding the mechanical properties of reactor structural materials in operation.		
		
		
				
		In summary, deformation localisation has been characterised in ion-irradiated Fe and Fe10Cr following nanoindentation. 
		Following irradiation, there is an increase in pile-up height, localisation of the pile-up topography, and formation of slip steps. Lattice rotation and strain fields surrounding the indent in the irradiated material are confined to a volume less than 13\% of that in the unirradiated materials. In practice, the localisation of deformation and the confinement of plastic zones means that a material is less able to absorb energy associated with deformation, leading to embrittlement. Despite significant differences in irradiation hardening, the irradiated Fe and Fe10Cr materials show very similar pile-up topography and deformation fields. The fact that Cr content does not seem to mitigate the loss of strain hardening capacity means that strategies other than varying the composition of steels are required to retain post-irradiation ductility, a key finding to guide the design of future reactor component materials.
		\section*{Declarations}
		\subsection*{Funding}
		This research used resources of the Advanced Photon Source, a U.S. Department of Energy (DOE) Office of Science User Facility operated for the DOE Office of Science by Argonne National Laboratory under Contract No. DE-AC02-06CH11357. The study was also supported by Laboratory Direct Research and Development (LDRD) funding from Argonne National Laboratory, provided by the Office of Science (OS), of the U.S. Department of Energy (DOE) under the same contract. The  authors  acknowledge  use  of  characterisation  facilities  within  the  David  Cockayne  Centre for  Electron  Microscopy,  Department  of  Materials,  University  of  Oxford,  alongside  financial support  provided  by  the  Henry  Royce  Institute (grant EP/R010145/1). KS acknowledges funding from the General Sir John Monash Foundation and the University of Oxford Department of Engineering Science. FH acknowledges funding from the European Research Council (ERC) under the European Union’s Horizon 2020 research and innovation programme (grant 714697). DEJA acknowledges funding from EPSRC (grant EP/P001645/1).
		
		\subsection*{Conflicts of interest}
		The authors have no relevant financial or non-financial interests to disclose.
		
		\subsection*{Data and code availability}
		All data, raw and processed, as well as the processing and plotting scripts are available at: \textit{A link will be provided after the review process and before publication.}
		
		\section*{Acknowledgements}
		The authors are grateful to R J Scales (Department of Materials, University of Oxford) for their help with AFM training.

\printbibliography

\end{document}


\begin{center}

\LARGE{\textbf{Supplementary File: Deformation localisation in ion-irradiated FeCr}}\\
\bigskip
\large
{Kay Song $^{\text{a}, \star}$, Dina Sheyfer $^{\text{b}}$, Wenjun Liu $^{\text{b}}$, Jonathan Z Tischler $^{\text{b}}$, Suchandrima Das $^{\text{c}}$, Kenichiro Mizohata $^{\text{d}}$, Hongbing Yu $^{\text{f}}$, David E J Armstrong $^{\text{e}}$, Felix Hofmann $^{\text{a}, \dagger}$}\\
\bigskip
\small{$^{\text{a}}$ Department of Engineering Science, University of Oxford, Parks Road, Oxford, OX1 3PJ, UK} \\
\small{$^{\text{b}}$ X-ray Science Division, Argonne National Laboratory, 9700 South Cass Avenue, Lemont, IL 60439, USA} \\
\small{$^{\text{c}}$ Department of Mechanical Engineering, University of Bristol, Queen's Building, University Walk, Bristol, BS8 1TR, UK} \\
\small{$^{\text{d}}$ Department of Physics, University of Helsinki, P.O. Box 43, 00014 Helsinki, Finland} \\
\small{$^{\text{e}}$ Department of Materials, University of Oxford, Parks Road, Oxford, OX1 3PH, UK}\\
\small{$^{\text{f}}$ Canadian Nuclear Laboratories, Chalk River, ON K0J 1J0, Canada} \\

\bigskip
\small{Corresponding author email: $^{\star}$kay.song@eng.ox.ac.uk; $\;^{\dagger}$felix.hofmann@eng.ox.ac.uk}
\end{center}

\baselineskip24pt

\renewcommand{\thefigure}{S-\arabic{figure}}
\section{AFM Deflection Error}

\begin{figure}[h!]
	\centering
	\includegraphics[width=0.65\textwidth]{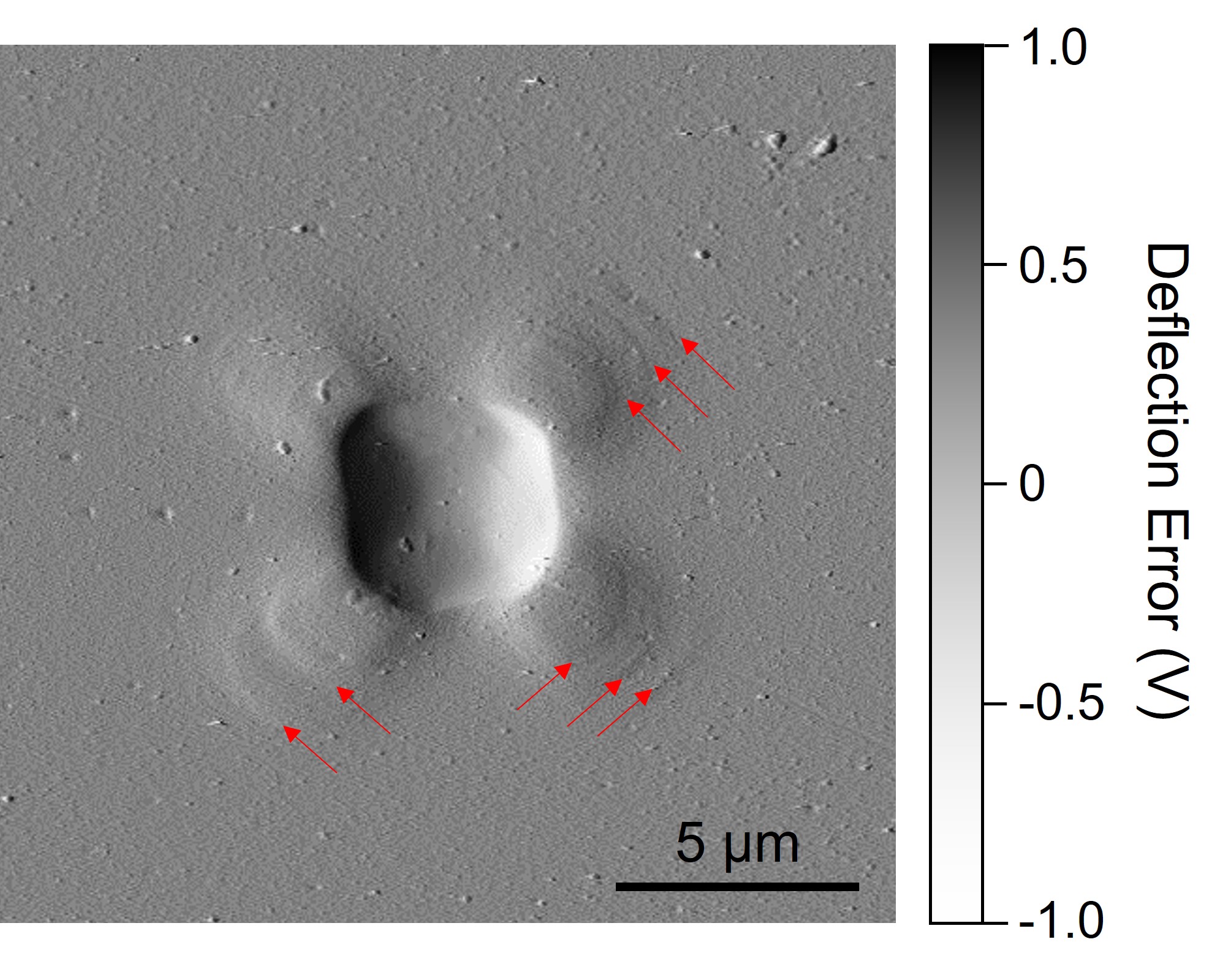}
	\caption{The deflection error of the scan surrounding a nanoindent on the irradiated Fe sample. Red arrows point to regions where slip steps can be easily identified.}
	\label{fig:grad}
\end{figure}
A plot of the deflection error, which is the difference between the measured deflection of the AFM probe and the set point, of the scan of a nanoindent on the irradiated Fe sample (Figure \ref{fig:grad}). From this plot, the slip steps can be more easily identified in the pile-up lobes. 

\section{DAXM Set-Up}
\begin{figure}[h!]
	\centering
	\includegraphics[width=0.5\textwidth]{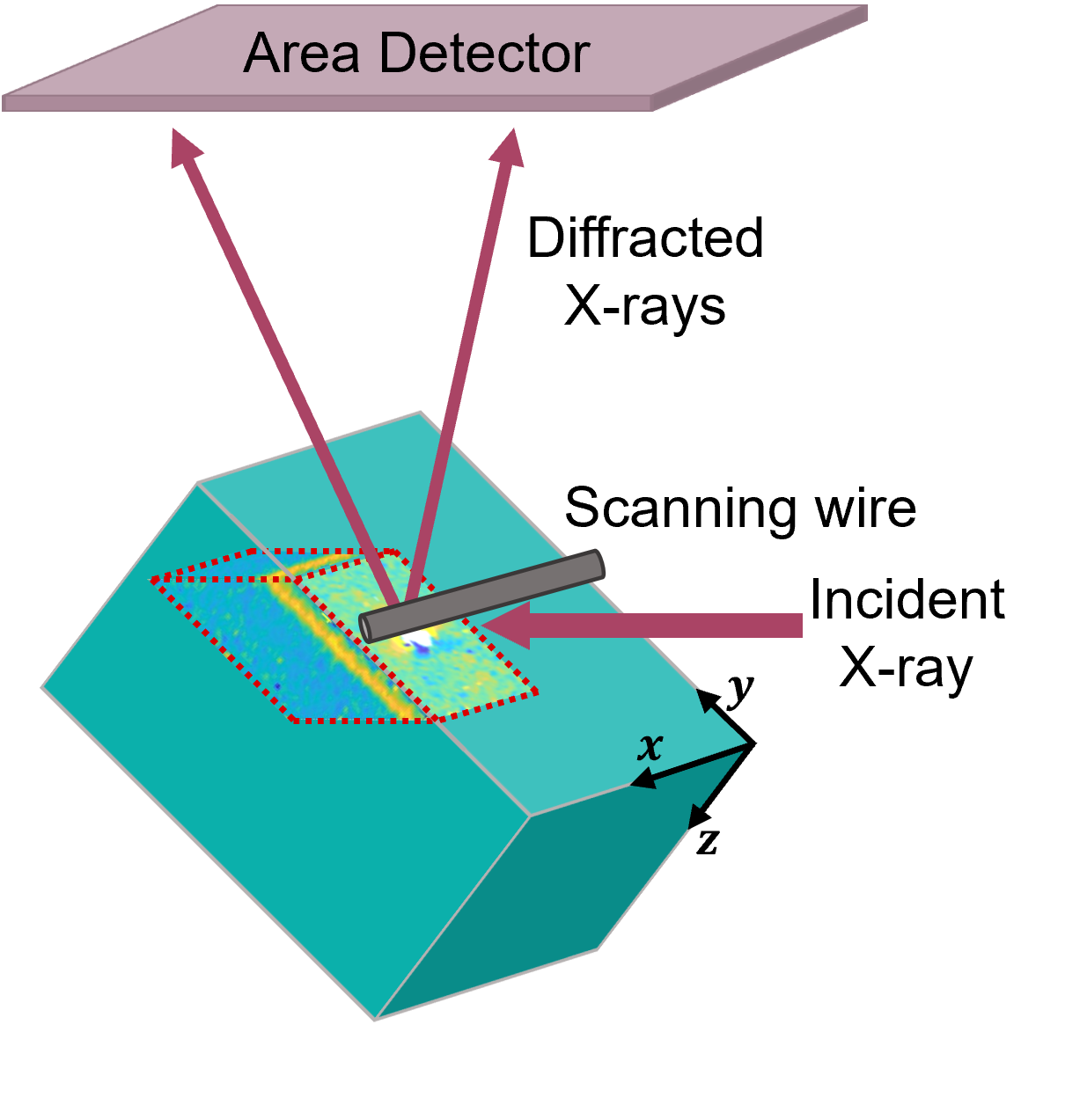}
	\caption{A schematic of the DAXM measurements showing the coordinates used in this study, the relative position of the indent, the scanned volume (dashed red lines), the bulk sample, and the beamline set-up. The example data shown here is from the Fe10Cr irradiated sample. The relative size of the scanned volume and the bulk sample is not to scale. }
	\label{fig:daxm_setup}
\end{figure}

The sample is mounted in a 45$^{\circ}$ reflection geometry. The diffraction patterns are captured by a 2D area detector (Perkin-Elmer, \#XRD 1621) positioned $\sim$511 mm above the sample. A platinum wire, of $\sim$100 \textmu m in diameter, is scanned parallel to the sample surface, perpendicular to the incoming beam. As it cuts through the diffracted beams, it blocks signal from certain depths in the sample. By comparing the recorded intensity associated with different wire positions, an intensity profile as a function of depth along the beam path can be calculated for each detector pixel. Further information on the DAXM technique can be found in \cite{Larson2002, Liu2004, Das2018a}.

We also note that with polychromatic X-ray Laue measurements, only the deviatoric strain components are measured since the there is no information on the radial positions of the diffraction peaks in reciprocal space \cite{Chung1999}. This is because the volumetric components of strain can only be determined if the the absolute lattice spacing is known from monochromatic X-ray Laue measurements. There are two ways to determine the full elastic strain tensor. The first is to obtain a monochromatic measurement of one diffraction peak to determine the absolute lattice plane spacing, then combine it with the deviatoric strain measurements from polychromatic X-ray Laue diffraction \cite{Hofmann2015, Robach2011}. The second method is to conduct energy scans to determine the absolute reciprocal space position of at least three non-collinear reflections \cite{Das2018a, Chung1999, Levine2015}.

Depth-resolved energy scans of the irradiated samples have been previously reported in \cite{Song2023}, but only for isolated points on the sample to determine the absolute change in lattice spacing due to ion irradiation. However, these energy scans were not performed for the 3D volume around the nanoindents in this study as the energy scan measurements take a long time and the deviatoric strains are sufficient for studying the deformation characteristics.

\section[Lattice Rotations and Strains in the $x-y$ Plane]{Lattice Rotations and Strains in the $\boldsymbol{x}-\boldsymbol{y}$ Plane}

\begin{figure}[h!]
	\centering
	\includegraphics[width=0.95\textwidth]{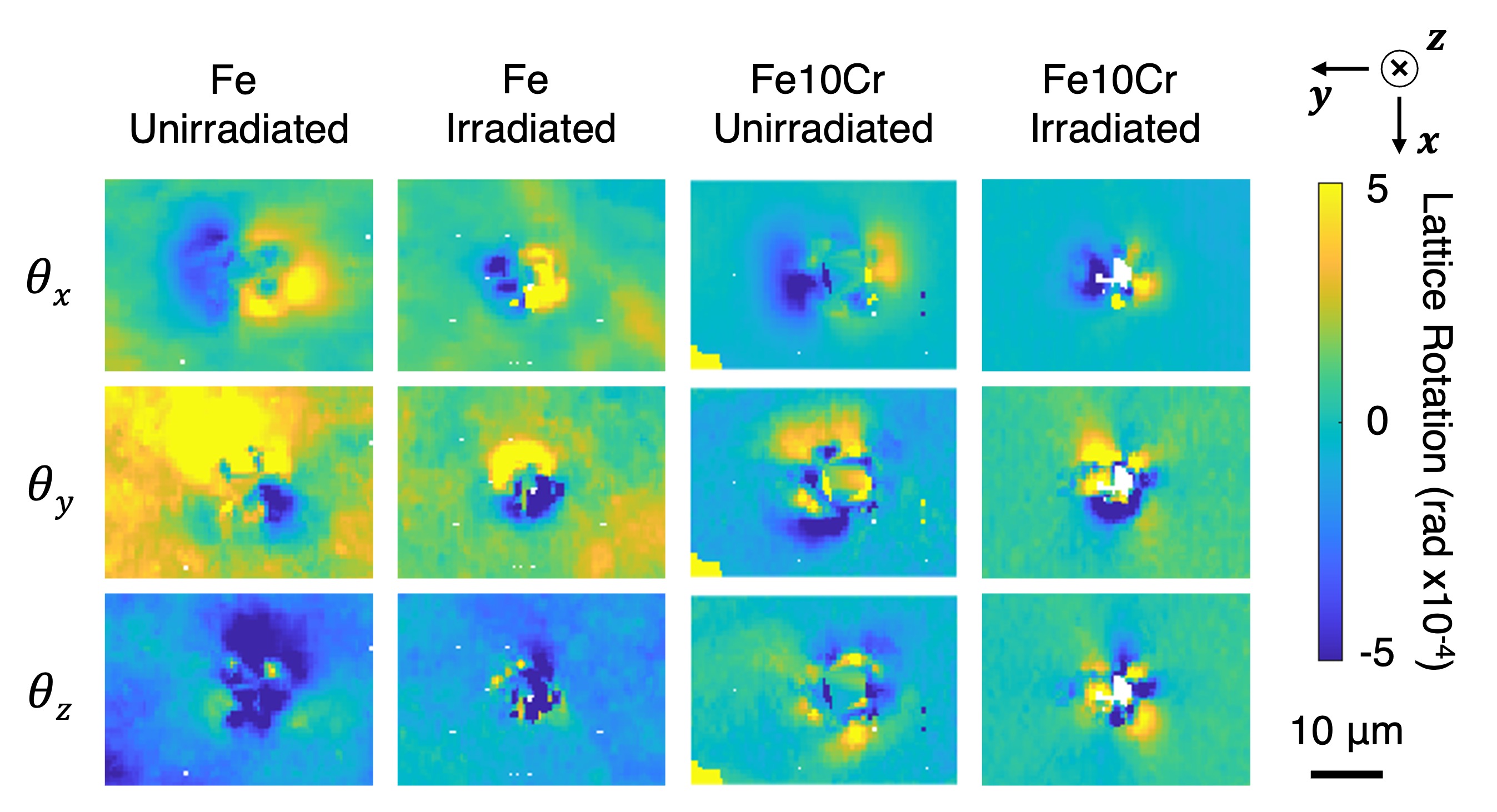}
	\caption{Lattice rotation in the $\boldsymbol{x}-\boldsymbol{y}$ plane at a depth of 2.5--3 \textmu m below the surface. }
	\label{fig:xyrot}
\end{figure}

\begin{figure}[h!]
	\centering
	\includegraphics[width=0.95\textwidth]{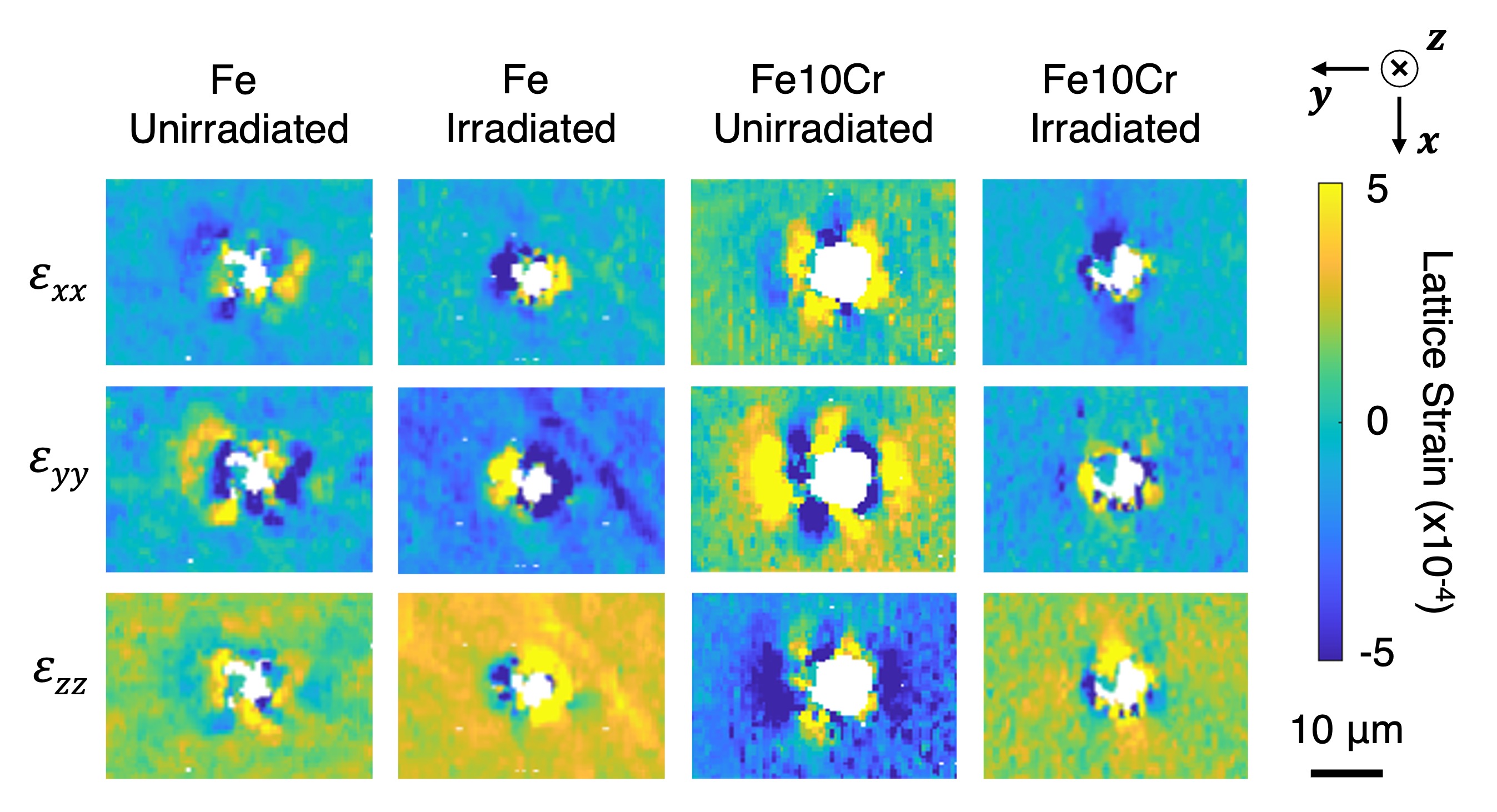}
	\caption{Lattice strains in the $\boldsymbol{x}-\boldsymbol{y}$ plane at a depth of 2.5--3 \textmu m below the surface. }
	\label{fig:xystrain}
\end{figure}

Lattice rotations and strains have been plotted in the $\boldsymbol{x}-\boldsymbol{y}$ plane, which is parallel to the bulk sample surface, at a slice taken from 2.5--3 \textmu m below the surface (Figures \ref{fig:xyrot} and \ref{fig:xystrain}). The spatial symmetries on the rotation and strain fields are as expected for indentation with a spherical indenter. 

Deformation localisation can be seen in both the lattice rotation and strain fields in the irradiated materials. In particular, the lattice strain field of the indent in the unirradiated Fe10Cr material has a very wide spatial distribution, up to 20 \textmu m from the indent centre (Figure \ref{fig:xystrain}). 
However, for the irradiated material the strain field is significantly more confined, only extending up to 10 \textmu m away from the indent centre. 

The distribution of large positive strain in the $\boldsymbol{z}$ direction for the irradiated Fe sample, even far away from the indent, is most likely due to the presence of the injected ions from the ion irradiation, as discussed in the main text and previously in \cite{Song2020, Song2023}.

\setlength{\baselineskip}{0pt} 
\newpage
\printbibliography